\documentclass{ws-p9-75x6-50}

\begin{document}

\title{Teleparallel Gravity: An Overview}

\author{V. C. de Andrade}

\address{D\'epartement d'Astrophysique Relativiste et de Cosmologie \\
Centre National de la Recherche Scientific (UMR 8629) \\
Observatoire de Paris, 92195 Meudon Cedex, France}

\author{L. C. T. Guillen and J. G. Pereira}

\address{Instituto de F\'{\i}sica Te\'orica,
Universidade Estadual Paulista\\
Rua Pamplona 145, 01405-900\, S\~ao Paulo, Brazil}

\maketitle

\abstracts{
The fundamentals of the teleparallel equivalent of general relativity are
presented, and its main properties described. In particular, the field
equations, the definition of an energy--momentum density for the
gravitational field, the teleparallel version of the equivalence principle,
and the dynamical role played by torsion as compared to the corresponding
role played by curvature in general relativity, are discussed in some
details.}

\section{Introduction}

Teleparallel gravity\cite{moller}$^-$\cite{mielke} corresponds to a gauge theory
for the translation group.\cite{paper1,hene} Due to the peculiar character of
translations, any gauge theory including these transformations will differ from the
usual internal gauge models in many ways, the most significant being the presence
of a tetrad field. On the other hand, a tetrad field can naturally be used to
define a linear Weitzenb\"ock connection, which is a connection presenting torsion,
but no curvature. A tetrad field can also naturally be used to define a
riemannian metric, in terms of which a Levi--Civita connection can be
constructed. As is well known, it is a connection presenting curvature,
but no torsion. Now, torsion and curvature are properties of a
connection,\cite{livro} and many different connections can be defined on
the same space.\cite{koba} Therefore one can say that the presence of a
nontrivial tetrad field in a gauge theory induces both a teleparallel and
a riemannian structures in spacetime. The first is related to the
Weitzenb\"ock,\cite{weitz} and the second to the Levi--Civita connection. Owing to
the universality of the gravitational interaction, it turns out to be possible to
link these geometrical structures to gravitation.

In the context of teleparallel gravity, curvature and torsion are able to
provide each one equivalent descriptions of the gravitational interaction.
Conceptual differences, however, show up. According to general relativity,
curvature is used to {\it geometrize} spacetime, and in this way
successfully describe the gravitational interaction. Teleparallelism, on the
other hand, attributes gravitation to torsion, but in this case torsion
accounts for gravitation not by geometrizing the interaction, but by acting
as a {\it force}. This means that, in the teleparallel equivalent of general
relativity, there are no geodesics, but force equations quite analogous to
the Lorentz force equation of electrodynamics.\cite{paper1} Thus, we can say
that the gravitational interaction can be described {\em alternatively} in
terms of curvature, as is usually done in general relativity, or in terms of
torsion, in which case we have the so called teleparallel gravity. Whether
gravitation requires a curved or a torsioned spacetime, therefore, turns
out to be a matter of convention.

In this paper, we are going to review the main features of teleparallel gravity. We
start in Sec.~2 where we introduce the fundamentals of the teleparallel equivalent of
general relativity, and discuss some of its main features. A detailed discussion of
the energy-momentum current for the gravitational field is made in Sec.~3. In Sec.~4
the second Bianchi identity of a gauge theory for the translation group is obtained,
from which we get the conservation law of the energy-momentum tensor for a general
matter (source) field. In Sec.~5 we study the motion of a spinless particle
in a gravitational field, and the gravitational analog of the Lorentz force
equation is obtained. Then, a discussion on the riemannian and teleparallel
versions of the equivalence principle is made. Finally in Sec.~6 we
draw the main conclusions of the paper.  

\section{Teleparallel Equivalent of General Relativity}

We use the Greek alphabet $(\mu, \nu, \rho, \dots = 0,1,2,3)$ to denote
indices related to spacetime (base space), and the Latin alphabet $(a,b,c,
\dots = 0,1,2,3)$ to denote indices related to the tangent space (fiber),
assumed to be a Minkowski space with the metric
$\eta_{ab}=\mbox{diag}(+1,-1,-1,-1)$.
A gauge transformation is defined as a local translation of the
tangent-space coordinates,
\begin{equation}
\delta x^{a} = \delta\alpha^{b}P_{b}x^{a} \; ,
\end{equation}
with $P_{a} = \partial /\partial x^a$ the translation generators, and
$\delta \alpha^{a}$ the corresponding infinitesimal parameters. Denoting the
gauge potentials by $A^{a}{}_{\mu}$, the gauge covariant derivative of a
general matter field $\Psi$ is\cite{paper1}
\be
{\mathcal D}_\mu \Psi = h^{a}{}_{\mu} \; \partial_{a} \Psi \; ,
\ee
where
\be
h^{a}{}_{\mu} = \partial_{\mu}x^{a} + c^{-2}A^{a}{}_{\mu}  
\label{2.17}
\ee
is a nontrivial tetrad field, with $c$ the speed of light. From the
covariance of
$D_{\mu} \Psi$, we obtain the transformation of the gauge potentials:
\begin{equation}
A^{a^{\prime}}{}_{\mu} = A^{a}{}_{\mu} - 
c^{2}\partial_{\mu}\delta\alpha^{a} \; .
\end{equation}
As usual in abelian gauge theories, the field strength is given by
\begin{equation}
F^{a}{}_{\mu \nu} = \partial_{\mu} A^{a}{}_{\nu}
- \partial_{\nu}A^{a}{}_{\mu} \; ,
\label{core}
\end{equation}
which satisfies the relation
\begin{equation}
[{\mathcal D}_{\mu}, {\mathcal D}_{\nu}] \Psi =
c^{-2} F^{a}{}_{\mu \nu} P_a \Psi \; .
\end{equation}
It is important to remark that, whereas the tangent space indices are
raised and lowered with the metric $\eta_{a b}$, the spacetime indices are
raised and lowered with the riemannian metric
\be
g_{\mu \nu} = \eta_{a b} h^a{}_{\mu} \, h^b{}_{\nu} \; .
\label{gmn}
\ee

A nontrivial tetrad field induces on spacetime a teleparallel structure
which is directly related to the presence of the gravitational field. In
fact, given a nontrivial tetrad, it is possible to define the so called
Weitzenb\"ock connection
\be
\Gamma^{\rho}{}_{\mu \nu} = h_{a}{}^{\rho}\partial_{\nu}h^{a}{}_{\mu} \;  ,
\label{carco}
\ee
which is a connection presenting torsion, but no curvature.\cite{livro} As
a natural consequence of this definition, the Weitzenb\"ock covariant derivative of
the tetrad field vanishes identically:
\be
\nabla_{\nu}h^a{}_{\mu} \equiv \partial_{\nu}h^a{}_{\mu} - 
\Gamma^{\theta}{}_{\mu \nu} \, h^a{}_{\theta} = 0 \; .
\label{cacd}
\ee
This is the absolute parallelism condition. The torsion of the Weitzenb\"ock
connection is
\be
T^{\rho}{}_{\mu \nu} = \Gamma^{\rho}{}_{\nu \mu} -
\Gamma^{\rho}{}_{\mu \nu} \; ,
\label{tor}
\ee
from which we see that the gravitational field strength is nothing but torsion
written in the tetrad basis:
\be
F^{a}{}_{\mu \nu} = c^{2}h^{a}{}_{\rho}T^{\rho}{}_{\mu \nu} \; .
\ee

A nontrivial tetrad field can also be used to define a torsionless linear
connection, the Levi-Civita connection of the metric (\ref{gmn}):
\be
{\stackrel{\circ}{\Gamma}}{}^{\sigma}{}_{\mu \nu} = \frac{1}{2} 
g^{\sigma \rho} \left[ \partial_{\mu} g_{\rho \nu} + \partial_{\nu}
g_{\rho \mu} - \partial_{\rho} g_{\mu \nu} \right] \; .
\label{lci}
\ee
The Weitzenb\"ock and the Levi--Civita connections are related by
\be
\Gamma^{\rho}{}_{\mu \nu} = 
{\stackrel{\circ}{\Gamma}}{}^{\rho} {}_{\mu \nu} + 
K^{\rho}{}_{\mu \nu} \; ,
\label{rela}
\ee
where
\be
K^{\rho}{}_{\mu \nu} = {\textstyle \frac{1}{2}} \left( T_{\mu}{}^{\rho}{}_{\nu}
+ T_{\nu}{}^{\rho}{}_{\mu} - T^{\rho}{}_{\mu \nu} \right)
\ee
is the contorsion tensor.

As already remarked, the curvature of the Weitzenb\"ock connection vanishes
identically:
\be
{R}^{\rho}{}_{\theta \mu \nu} = \partial_\mu
{\Gamma}^{\rho}{}_{\theta \nu} + {\Gamma}^{\rho}{}_{\sigma
\mu} \; {\Gamma}^{\sigma}{}_{\theta \nu} - (\mu  \leftrightarrow \nu)
\equiv 0 \; .
\label{r}
\ee
Substituting ${\Gamma}^{\rho}{}_{\mu \nu}$ as given by
Eq.(\ref{rela}), we get
\be
{R}^{\rho}{}_{\theta \mu \nu} =
{\stackrel{\circ}{R}}{}^{\rho}{}_{\theta \mu \nu} +
Q^{\rho}{}_{\theta \mu \nu} \equiv 0 \; ,
\label{relar}
\ee
where ${\stackrel{\circ}{R}}{}^{\theta}{}_{\rho \mu \nu}$ is the
curvature of the Levi--Civita connection, and
\be
Q^{\rho} {}_{\theta \mu \nu} = {D}_{\mu}{}{K}^{\rho}{}_{\theta \nu} - 
{D}_{\nu}{}{K}^{\rho}{}_{\theta \mu} + {K}^{\sigma}{}_{\theta \nu}
\; {K}^{\rho}{}_{\sigma \mu} - {K}^{\sigma}{}_{\theta \mu} \;
{K}^{\rho}{}_{\sigma \nu}
\label{qdk}
\ee
is a tensor written in terms of the Weitzenb\"ock connection only. Here,
$D_\mu$ is the teleparallel covariant derivative, which is nothing but the
Levi-Civita covariant derivative of general relativity rephrased in terms
of the Weitzenb\"ock connection.~\cite{vector} Acting on a spacetime vector
$V^\mu$, for  example, its explicit form is
\be
D_\rho \, V^\mu \equiv \partial_\rho V^\mu +
\left( \Gamma^\mu{}_{\lambda \rho} - K^\mu{}_{\lambda \rho} \right) V^\lambda \; .
\label{tcd}
\ee

Equation (\ref{relar}) has an interesting interpretation: the contribution
${\stackrel{\circ}{R}}{}^{\rho}{}_{\theta \mu \nu}$ coming from the
Levi--Civita connection compensates exactly the contribution
$Q^{\rho}{}_{\theta \mu \nu}$ coming from the Weitzenb\"ock connection, yielding
an identically zero curvature tensor ${R}^{\rho}{}_{\theta \mu \nu}$. This is a
constraint satisfied by the Levi--Civita and Weitzenb\"ock connections, and is the
fulcrum of the equivalence between the riemannian and the teleparallel descriptions
of gravitation.

The gauge gravitational field Lagrangian is given by\cite{paper1}
\be
{\cal L}_G = 
\frac{h c^{4}}{16 \pi G} \; S^{\rho \mu \nu} \; T_{\rho \mu \nu} \; ,
\label{gala}
\ee
where $h = {\rm det}(h^{a}{}_{\mu})$, and
\[
S^{\rho \mu \nu} = - S^{\rho \nu \mu} \equiv {\textstyle \frac{1}{2}} 
\left[ K^{\mu \nu \rho} - g^{\rho \nu} \; T^{\theta \mu}{}_{\theta} + g^{\rho \mu}
\; T^{\theta \nu}{}_{\theta} \right] 
\]
is a tensor written in terms of the Weitzenb\"ock connection only. As usual in
gauge theories, it is quadratic in the field strength. By using relation
(\ref{rela}), this lagrangian can be rewritten in terms of the Levi-Civita
connection only. Up to a total divergence, the result is the Hilbert--Einstein
Lagrangian of general relativity
\be
{\cal L} = - \frac{c^4}{16 \pi G} \;  \sqrt{-g} \, {\stackrel{\circ}{R}} \; ,
\ee
where the identification $h = \sqrt{-g}$ has been made.

By performing variations in relation to the gauge field $A_a{}^\rho$, we obtain
from the gauge lagrangian ${\cal L}_G$ the teleparallel version of the
gravitational field equation,
\be
\partial_\sigma(h S_a{}^{\sigma \rho}) - 
\frac{4 \pi G}{c^4} \, (h j_{a}{}^{\rho}) = 0 \; ,
\label{tfe1}
\ee
where $S_a{}^{\sigma \rho} \equiv h_{a}{}^{\lambda}
S_{\lambda}{}^{\sigma \rho}$. Analogously to the Yang-Mills theories, 
\be
h j_{a}{}^{\rho} \equiv \frac{\partial {\cal L}_G}{\partial h^a{}_{\rho}} = -
\frac{c^{4}}{4 \pi G} \, h h_a{}^{\lambda} S_{\mu}{}^{\nu \rho} 
T^\mu{}_{\nu \lambda} + h_a{}^{\rho} {\cal L}_G
\label{ptem1} 
\ee
stands for the gauge current, which in this case represents the energy and
momentum of the gravitational field.\cite{gemt} The term $(h S_a{}^{\sigma \rho})$
is called {\it superpotential} in the sense that its derivative yields the gauge
current $(h j_{a}{}^{\rho})$. Due to the anti-symmetry of $S_a{}^{\sigma \rho}$
in the last two indices, $(h j_{a}{}^{\rho})$ is conserved as a consequence of the
field equation:
\be
\partial_\rho (h j_a{}^\rho) = 0 \; .
\label{conser1}
\ee
Making use of the identity
\be
\partial_\rho h \equiv h {\Gamma}^{\nu}{}_{\nu \rho} =
h \left( {\Gamma}^{\nu}{}_{\rho \nu} - K^{\nu}{}_{\rho \nu} \right) \; ,
\label{id1}
\ee
this conservation law can alternatively be written in the form
\be
D_\rho \, j_a{}^\rho \equiv \partial_\rho j_a{}^\rho +
\left( \Gamma^\rho{}_{\lambda \rho} - K^\rho{}_{\lambda \rho} \right) 
j_a{}^\lambda = 0 \; ,
\label{conser2}
\ee
with $D_\rho$ denoting the teleparallel version of the covariant derivative,
which is nothing but the Levi-Civita covariant derivative of general relativity
rephrased in terms of the Weitzenb\"ock connection.\cite{vector}

\section{Gravitational Energy-Momentum Current} 

Now comes an important point. As can be easily checked, the current $j_a{}^\rho$
transforms covariantly under a general spacetime coordinate transformation, is
invariant under local (gauge) translation of the tangent-space coordinates, and
transforms covariantly under a global tangent--space Lorentz transformation.
This means that $j_a{}^\rho$, despite not covariant under a local Lorentz
transformation, is a true spacetime and gauge tensor.\cite{gemt}

Let us now proceed further and find out the relation between the above gauge
approach and general relativity. By using Eq.~(\ref{carco}) to express
$\partial_\rho h_a{}^\lambda$, the field equation (\ref{tfe1}) can be rewritten in
a purely spacetime form,
\be
\partial_\sigma(h S_\lambda{}^{\sigma \rho}) - 
\frac{4 \pi G}{c^4} \, (h t_{\lambda}{}^{\rho}) = 0 \; ,
\label{tfe2}
\ee
where now 
\be
h t_{\lambda}{}^{\rho} =
\frac{c^{4} h}{4 \pi G} \, \Gamma^{\mu}{}_{\nu \lambda} S_{\mu}{}^{\nu \rho}
+ \delta_\lambda{}^{\rho} {\cal L}_G
\label{ptem2} 
\ee
stands for the canonical energy-momentum pseudotensor of the gravitational
field.\cite{shirafuji96} Despite not apparent, Eq.~(\ref{tfe2}) is symmetric in
$(\lambda \rho)$. Furthermore, by using Eq.~(\ref{rela}), it can be rewritten in
terms of the Levi-Civita connection only. As expected, due to the equivalence
between the corresponding Lagrangians, it is the same as Einstein's equation:
\be
\frac{h}{2} \left[{\stackrel{\circ}{R}}_{\mu \nu} -
\frac{1}{2} \, g_{\mu \nu}
{\stackrel{\circ}{R}} \right] = 0 \; .
\ee  

It is important to notice that the canonical energy-momentum pseudotensor
$t_{\lambda}{}^{\rho}$ is not simply the gauge current $j_a{}^\rho$ with the
algebraic index ``$a$'' changed to the spacetime index ``$\lambda$''. It
incorporates also an extra term coming from the derivative term of
Eq.~(\ref{tfe1}):
\be
t_\lambda{}^\rho = h^a{}_\lambda \, j_a{}^\rho +
\frac{c^{4}}{4 \pi G} \, \Gamma^{\mu}{}_{\lambda \nu} S_{\mu}{}^{\nu \rho} \; .
\label{ptem3}
\ee
We see thus clearly the origin of the connection-term which transforms the gauge
current $j_a{}^\rho$ into the energy-momentum pseudotensor $t_\lambda{}^\rho$.
Through the same mechanism, it is possible to appropriately exchange further terms
between the derivative and the current terms of the field equation (\ref{tfe2}),
giving rise to different definitions for the energy-momentum pseudotensor, each
one connected to a different {\it superpotential} $(h S_\lambda{}^{\rho \sigma})$.

Like the gauge current $(h j_a{}^\rho)$, the pseudotensor $(h t_\lambda{}^\rho)$
is conserved as a consequence of the field equation:
\be
\partial_\rho (h t_\lambda{}^\rho) = 0 \; .
\label{conser3}
\ee
However, in contrast
to what occurs with $j_a{}^\rho$, due to the pseudotensor character of
$t_\lambda{}^\rho$, this conservation law can not be rewritten with a covariant
derivative.

Because of its simplicity and transparency, the teleparallel approach to
gravitation seems to be much more appropriate than general relativity to deal with
the energy problem of the gravitational field. In fact, M{\o}ller already noticed
a long time ago that a satisfactory solution to the problem of the energy
distribution in a gravitational field could be obtained in the framework of a
tetrad theory. In our notation, his expression for the gravitational
energy-momentum density is\cite{moller}
\be
h t_\lambda{}^\rho = \frac{\partial {\cal L}}{\partial \partial_\rho h^a{}_\mu} \;
\partial_\lambda h^a{}_\mu + \delta_\lambda{}^\rho \, {\cal L} \; ,
\ee
which is nothing but the usual canonical energy-mo\-men\-tum density yielded by
Noether's theorem. Using for ${\cal L}$ the gauge Lagrangian (\ref{gala}), it is
an easy task to verify that M{\o}ller's expression coincides exactly with the
teleparallel energy-momentum density appearing in the field equation
(\ref{tfe2}-\ref{ptem2}). Since $j_a{}^\rho$ is a true spacetime tensor, whereas
$t_\lambda{}^\rho$ is not, we can say that the gauge current $j_a{}^\rho$ is an
improved version of the M{\o}ller's energy-momentum density $t_\lambda{}^\rho$.
Mathematically, they can be obtained from each other by using the relation
(\ref{ptem3}). It should be remarked, however, that both of them transform
covariantly only under {\it global} tangent-space Lorentz transformations. The lack
of a {\it local} Lorentz covariance can be considered as the teleparallel
manifestation of the pseudotensor character of the gravitational energy-momentum
density in general relativity.\cite{gemt}

\section{Bianchi Identity and Matter Energy-Momentum Conservation}

As is well known, the second Bianchi identity of general relativity is
\be
{\stackrel{\circ}{\nabla}}_\sigma {\stackrel{\circ}{R}}_{\lambda \rho \mu \nu} +
{\stackrel{\circ}{\nabla}}_\nu {\stackrel{\circ}{R}}_{\lambda \rho \sigma \mu} +
{\stackrel{\circ}{\nabla}}_\mu {\stackrel{\circ}{R}}_{\lambda \rho \nu \sigma} = 0
\; ,
\label{birg2}
\ee
where ${\stackrel{\circ}{\nabla}}_\mu$ is the usual Levi-Civita covariant
derivative. Its contracted form is
\be
{\stackrel{\circ}{\nabla}}_\mu \left[ {\stackrel{\circ}{R}}{}^\mu{}_\nu -
\textstyle{\frac{1}{2}} \delta^\mu{}_\nu {\stackrel{\circ}{R}} \right] = 0 \; .
\ee
By using Eq.(\ref{relar}), after a tedious but straightforward calculation, it is
possible to rewrite it in terms of the Weitzenb\"ock connection only. The result is
\be
D_\rho \left[ \partial_\sigma(h S_\lambda{}^{\sigma \rho}) - 
\frac{4 \pi G}{c^4} \, (h t_{\lambda}{}^{\rho}) \right] = 0 \; ,
\label{bi3}
\ee
where $D_\rho$ is the teleparallel covariant derivative, defined in
Eq.(\ref{conser2}). This is the second Bianchi identity of the teleparallel
equivalent of general relativity. It says that the teleparallel covariant
derivative of the sourceless field equation (\ref{tfe2}) vanishes identically.

In the presence of a general matter field, the teleparallel field equation
(\ref{tfe2}) becomes
\be
\partial_\sigma(h S_\lambda{}^{\sigma \rho}) - 
\frac{4 \pi G}{c^4} \, (h t_{\lambda}{}^{\rho}) =
\frac{4 \pi G}{c^4} \, (h {\cal T}_{\lambda}{}^{\rho}) \; ,
\label{tfe3}
\ee
with ${\cal T}_{\lambda}{}^{\rho}$ the matter energy-momentum tensor. As a
consequence of the Bianchi identity (\ref{bi3}), and using (\ref{id1}), we obtain
\be
D_\rho {\cal T}_\lambda{}^\rho = 0 \; .
\label{telecon}
\ee
This is the conservation law of matter energy-momentum tensor. Therefore, we see
that in teleparallel gravity, it is not the Weitzenb\"ock covariant derivative
$\nabla_\mu$, but the teleparallel covariant derivative (\ref{conser2}) that
yields the correct conservation law for the energy-momentum tensors of matter
fields. It should be remarked that (\ref{telecon}) is the unique law compatible
with the corresponding conservation law of general relativity,
\be
{\stackrel{\circ}{\nabla}}{}_\mu {\cal T}^\mu{}_\rho \equiv 
\partial_\mu {\cal T}^\mu{}_\rho +
{\stackrel{\circ}{\Gamma}}{}^{\mu}{}_{\lambda \mu} {\cal T}^\lambda{}_\rho -
{\stackrel{\circ}{\Gamma}}{}^{\lambda}{}_{\rho \mu} {\cal T}^\mu{}_\lambda = 0 \; .
\label{grcon}
\ee
as can easily be verified by using Eq.(\ref{rela}).

\section{Geodesics Versus Force Equation}

In the framework of the teleparallel description of gravitation, the action
describing a particle of mass $m$ submitted to a gravitational ($A^a{}_{\mu}$)
field is\cite{paper1}
\begin{equation}
S=\int_{a}^{b}\left[ - m \, c \, d\sigma - \frac{m}{c} \, A^{a}{}_{\mu} \, 
u_{a} \, dx^{\mu} \right] \; ,
\label{3.1}
\end{equation}
where $d\sigma=(\eta_{a b} dx^a dx^b)^{1/2}$ is the invariant tangent-space
interval, and $u_a = {d x_a}/{d \sigma}$ is the tangent-space four-velocity. The
corresponding equation of motion is
\begin{equation}
c^{2} h^{a}{}_{\rho } \frac{du_{a}}{ds} = F^{a}{}_{\rho \mu }
u_{a}u^{\mu} \; ,
\label{3.2}
\end{equation}
where $ds=(g_{\mu \nu} dx^\mu dx^\nu)^{1/2}$ is the invariant spacetime interval,
and $u^\mu = {d x^\mu}/{d s}$ is the spacetime four velocity. It is important to
remark that, by using the relations
\[
h_a{}^\mu u^a u_\mu = 1 \; ,
\]
and
\[
\frac{\partial x^\mu}{\partial x^a} \, u^a \, u_\mu = \frac{d s}{d \sigma} \; ,
\]
the action (\ref{3.1}) reduces to its general relativity version
\[
S = - \int_{a}^{b} m c ds \; .
\]
In this case, the interaction of the particle with the gravitational field is
generated by the presence of the metric tensor $g_{\mu \nu}$ in 
$ds$, or alternatively in the constraint $u^2=g_{\mu \nu} u^\mu \, u^\nu=1$
if one opts for using a lagrangian formalism.

Now, by transforming algebra into spacetime indices, the equation of motion
(\ref{3.2}) can be rewritten alternatively in terms of
magnitudes related to the Weitzenb\"ock\cite{weitz} or to the Riemann
spacetime, giving rise respectively to the teleparallel and the
metric equations of motion. In fact, by using the relation
\be
h^{a}{}_{\mu} \frac{d u_{a}}{d s} = \omega_\mu \equiv
\frac{d u_{\mu}}{d s} - \Gamma_{\theta \mu \nu} u^{\theta}
\, u^{\nu} \; ,
\ee
where $\omega_\mu$ is the spacetime particle four--acceleration, it reduces to
\be
\frac{d u_\mu}{d s} - \Gamma_{\theta \mu \nu} \; u^\theta \;
u^\nu =  T_{\theta \mu \nu} \; u^\theta \; u^\nu \; .
\label{geode}
\ee
The left--hand side of this equation is the Weitzenb\"ock covariant derivative of
$u_\mu$ along the world line of the particle. The presence of the torsion tensor
on its right--hand side shows that it plays the role of an external force.
Substituting Eq.(\ref{tor}), it becomes
\be
\frac{d u_\mu}{d s} - \Gamma_{\theta \nu \mu} \; u^\theta \;
u^\nu = 0 \; .
\label{geodeflat}
\ee
It is important to remark that, as $\Gamma_{\theta \nu \mu}$ is not symmetric in
the last two indices, this is not a geodesic equation. This means that the
trajectories followed by spinless particles are not geodesics of the  induced
Weitzenb\"ock spacetime. In a locally inertial coordinate system,
$\partial_\mu g_{\theta \nu} = 0$, and the Weitzenb\"ock connection
$\Gamma_{\theta \nu \mu}$ becomes skew--symmetric in the first two indices. In
this coordinate system, therefore, owing to the symmetry of  $u^\theta \; u^\nu$,
the force equation (\ref{geodeflat}) becomes the equation of motion of a free
particle. This is the teleparallel version of the (weak) equivalence principle.

We transform again algebra into spacetime indices, but now in such a way to get
the force equation (\ref{3.2}) written in terms of the Levi--Civita connection
only.  Following the same steps used earlier, we get
\be
\frac{d u_\mu}{d s} - \Gamma_{\theta \mu \nu} \; u^\theta \; u^\nu = 
T_{\theta \mu \nu} \; u^\theta \; u^\nu \; .
\label{geode5}
\ee 
Then, by taking into account the symmetry of $u^\theta \; u^\nu$
under the exchange $(\theta \leftrightarrow \nu)$, we can rewrite
it as
\be
\frac{d u_\mu}{d s} - \Gamma_{\theta \mu \nu} \; u^\theta \;
u^\nu =  K_{\mu \theta \nu} \; u^\theta \; u^\nu \; .
\label{geode2}
\ee
Noticing that $K_{\mu \theta \nu}$ is skew--symmetric in the
first two indices, and using Eq.(\ref{rela}) to express
$(K_{\theta \mu \nu} - \Gamma_{\theta \mu \nu})$,
Eq.(\ref{geode2}) becomes
\be
\frac{d u_\mu}{d s} - {\stackrel{\circ}{\Gamma}}{}_{\theta \mu
\nu} \;  u^\theta \; u^\nu = 0 \; .
\label{geo2}
\ee
This is precisely the geodesic equation of general relativity, which means that
the trajectories followed by spinless particles  are geodesics of the induced
Riemann spacetime.  According to this description, therefore, the only effect of
the  gravitational field is to induce a {\em curvature} in spacetime, which will
then be the responsible for determining the trajectory of the particle. In a
locally inertial coordinate system, the first derivative of the metric tensor
vanishes, the Levi--Civita connection vanishes as well, and the geodesic equation
(\ref{geo2}) becomes the equation of motion of a free particle. This is the
usual version of the (weak) equivalence principle as formulated in general
relativity.\cite{weinberg}

Notice the difference in the index contractions between the connections and the
four--velocities in equations (\ref{geodeflat}) and (\ref{geo2}). This difference
is the responsible  for the different characters of these equations: the first is
a force equation written in the underlying Weitzenb\"ock  spacetime, and the
second is a true geodesic equation written in the induced Riemann spacetime. Now,
as both equations are deduced from the same force equation (\ref{3.2}), they  must
be equivalent ways of describing the same physical trajectory. In fact, it is
easy to see that any one of them can be obtained from the other by using the
relation (\ref{rela}).

\section{Conclusions}

In general relativity, the presence of a gravitational field is expressed by the
torsionless Levi--Civita metric--connection, whose curvature determines the
intensity of the gravitational field. On the other hand, in the teleparallel
description of gravitation, the presence of a gravitational field is expressed by
the flat Weitzenb\"ock connection, whose torsion is now the entity responsible for
determining the intensity of the gravitational field. The gravitational
interaction, therefore, can be described {\em either}, in terms of curvature, or
in terms of torsion. Whether gravitation requires a curved or a torsioned
spacetime, therefore, is a matter of convention.

An important point of the teleparallel equivalent of general relativity is that it
allows for the definition of an energy-momentum gauge current $j_a{}^\rho$
for the gravitational field which is covariant under a spacetime general
coordinate transformation, and transforms covariantly under a global
tangent-space Lorentz transformation. This means essentially that $j_a{}^\rho$ is
a true spacetime tensor, but not a tangent--space tensor. Then, by rewriting the
gauge field equation in a purely spacetime form, it becomes Einstein's equation,
and the gauge current $j_a{}^\rho$ reduces to the canonical energy-momentum
pseudotensor of the gravitational field. Teleparallel gravity, therefore, seems to
provide a more appropriate environment to deal with the energy problem since in the
ordinary context of general relativity, the energy-momentum density for the
gravitational field will always be represented by a pseudotensor.

Like in general relativity, the conservation law of the energy-momentum tensor
of matter (source) fields can be obtained from the Bianchi identities. In the
case of teleparallel gravity, the energy-momentum tensor turns out to be
conserved with the teleparallel covariant derivative, which is the
Levi-Civita covariant derivative of general relativity rephrased in terms of the
Weitzenb\"ock connection.\cite{vector}

Finally, we have succeeded in obtaining a gravitational analog of the Lorentz
force equation, which is an equation written in the underlying Weitzenb\"ock
spacetime.  According to this approach, the trajectory of the particle is
described in the very same way the Lorentz force describes the trajectory of a
charged particle in the presence of an electromagnetic field, with torsion playing
the role of force. When rewritten in terms of magnitudes related to the metric
structure, it becomes the geodesic equation of general relativity, which is an
equation written in the underlying Riemann spacetime. As both equations are
deduced from the same force equation, they  must be equivalent ways of describing
the same physical trajectory.

\section*{Acknowledgments}

The authors would like to thank R. Aldrovandi for useful discussions. They would
like also to thank FAPESP-Brazil, CAPES-Brazil and CNPq--Brazil for financial
support.


\begin{thebibliography}{99}

\bibitem{moller}
C. M{\o}ller, K. Dan. Vidensk. Selsk. Mat. Fys. Skr.{\bf 1}, No.
10 (1961).

\bibitem{pelle}
C. Pellegrini and J. Plebanski, K. Dan. Vidensk. Selsk. Mat. Fys.
Skr. {\bf 2}, No. 2 (1962).

\bibitem{haya}
K. Hayashi and T. Nakano, Prog. Theor. Phys. {\bf 38}, 491 (1967).

\bibitem{hayshi}
K. Hayashi and T. Shirafuji, Phys. Rev. D {\bf 19}, 3524 (1979).

\bibitem{hehl}
F. W. Hehl, in {\em Cosmology and Gravitation}, ed. by P. G.
Bergmann and  V. de Sabbata (Plenum, New York, 1980).

\bibitem{kopc}
W. Kopczy\'nski, J. Phys. A {\bf 15}, 493 (1982).

\bibitem{azeredo}
R. de Azeredo Campos and C. G. Oliveira, Nuovo Cimento B {\bf 74},
83 (1983).

\bibitem{nitsch1}
F. M\"uller--Hoissen and J. Nitsch, Gen. Rel. Grav. {\bf 17}, 747
(1985).

\bibitem{mielke}
E. W. Mielke, Ann. Phys. (NY) {\bf 219}, 78 (1992).

\bibitem{paper1}
V. C. de Andrade and J. G. Pereira, Phys. Rev D {\bf 56}, 4689 (1997).

\bibitem{hene}
For a general review on the gauge approach to gravity, see F. W. Hehl, J. D.
McCrea, E. W. Mielke and Y. Ne'eman, Phys. Rep.  {\bf 258}, 1 (1995).

\bibitem{livro}
R. Aldrovandi and J. G. Pereira, {\em An Introduction to Geometrical
Physics} (World Scientific, Singapore, 1995).

\bibitem{koba}
S. Kobayashi, and K. Nomizu, {\em Foundations of Differential Geometry}
(Intersciense, New YorkK, 1963).

\bibitem{weitz}
R. Weitzenb\"ock, {\em Invariantentheorie} (Noordhoff, Gronningen,
1923).

\bibitem{vector}
V. C. de Andrade and J. G. Pereira, Int. J. Mod. Phys. D {\bf 8}, 141 (1999).

\bibitem{gemt}
V. C. de Andrade, L. C. T. Guillen and J. G. Pereira,
Phys. Rev. Lett. {\bf 84}, 4533 (2000).

\bibitem{shirafuji96}
T. Shirafuji, G. L. Nashed and Y. Kobayashi, Prog. Theor. Phys.
{\bf 96}, 933 (1996);

\bibitem{weinberg}
See, for example, S. Weinberg, {\em Gravitation and Cosmology} 
(Wiley, New York, 1972).

\end{thebibliography}
\end{document}